\documentclass{cimento}
\usepackage[]{graphicx}
\usepackage{subfigure}
\usepackage[USenglish]{babel}

\def\bef{\begin{figure}}
\def\eef{\end{figure}}

\newcommand{\be}[1]{\begin{equation}\label{#1}}
\newcommand{\beq}{\begin{equation}}
\newcommand{\ee}{\end{equation}}
\newcommand{\beqn}[1]{\begin{eqnarray}\label{#1}}
\newcommand{\eeqn}{\end{eqnarray}}
\newcommand{\bd}{\begin{displaymath}}
\newcommand{\ed}{\end{displaymath}}
\newcommand{\mat}[4]{\left(\begin{array}{cc}{#1}&{#2}\\{#3}&{#4}
\end{array}\right)}

\def\lsim{\raise0.3ex\hbox{$\;<$\kern-0.75em\raise-1.1ex
e\hbox{$\sim\;$}}}
\def\gsim{\raise0.3ex\hbox{$\;>$\kern-0.75em\raise-1.1ex
\hbox{$\sim\;$}}}
\def\simlt{\mathrel{\lower2.5pt\vbox{\lineskip=0pt\baselineskip=0pt
           \hbox{$<$}\hbox{$\sim$}}}}
\def\simgt{\mathrel{\lower2.5pt\vbox{\lineskip=0pt\baselineskip=0pt
           \hbox{$>$}\hbox{$\sim$}}}}
\def\unity{{\hbox{1\kern-.8mm l}}}

\renewcommand{\to}{\rightarrow}

\def\cD{{\cal D}}
\def\cF{{\cal F}}
\def\cE{{\cal E}}

\def\cL{{\cal L}}
\def\cN{{\cal N}}

\def\cQ{{\cal Q}}

\def\cU{{\cal U}}
\def\lsim{\mathrel{\mathop  {\hbox{\lower0.5ex\hbox{$\sim$}
\kern-0.8em\lower-0.7ex\hbox{$<$}}}}}
\def\gsim{\mathrel{\mathop  {\hbox{\lower0.5ex\hbox{$\sim$}
\kern-0.8em\lower-0.7ex\hbox{$>$}}}}}

\title{Dark Matter and IceCube Neutrinos}
\author{R.~Biondi}
\instlist{\inst{} Dipartimento di Scienze Fisiche e Chimiche, Universit\`a di L'Aquila, 67100  L'Aquila (AQ), Italy
  \inst{} INFN, Laboratori Nazionali del Gran Sasso (LNGS), 67100 Assergi (AQ), Italy.}

\PACSes{\PACSit{14.60.Pq}{Neutrino mass and mixing}
\PACSit{14.60.St}{Non-standard-model neutrinos, right-handed neutrinos, etc.}}

\begin{document}

\maketitle

\begin{abstract}
We show that the excess of high energy neutrinos observed by the IceCube collaboration 
at energies above 100 TeV  
might originate from baryon number violating decays of heavy shadow  baryons from mirror sector,
which in turn constitute Dark Matter. Due to tiny mixing 
between mirror and ordinary neutrinos, it is possible to 
explain the specific features of the IceCube events spectrum. 
\end{abstract}

Discovery of high-energy neutrino events by the IceCube Collaboration 
\cite{IC} opened  a new era of experimental high-energy neutrino astrophysics. 
Recently the collaboration published the three year data collected 
between 2010 and 2013, containing 37 candidate in the energy range 
from 30 TeV to 2 PeV. While the number of events with energies below 100 TeV or so 
is compatible with the background expectations, there is an evident excess of the events 
with $E > 100$~TeV or so. On the other hand, no events were observed 
in the gap between 400 TeV and 1 PeV, but  
three most energetic shower events emerged 
at the end of the spectrum with energies 
1.04 PeV,  1.14 PeV and 2.0 PeV where the atmospheric background is practically vanishing.  
On the other hand,  the spectrum is apparently cut off at energies larger than about 2 PeV. 
The gap in the energy spectrum is difficult to explain in known models 
of high-energy neutrinos of astrophysical origin. 

The model that may explain such a spectrum was suggested in ref. \cite{BBDG}.  
Here we present  main guidelines of this model.  
Nowadays it becomes popular the idea that dark matter of the universe 
can emerge from a parallel gauge sector,  
with particles and interactions sharing many similarities with ordinary particle sector.  
Such a shadow sector would contain particles like quarks which form 
composite baryons, as well as leptons and neutrinos 
which are all sterile for ordinary gauge interactions.  
In particularly interesting example is represented by  
so-called mirror world \cite{Berezhiani:2003xm}, 
which has the content of particles and interactions exactly identical to that 
of ordinary sector, 
with the same gauge and Yukawa coupling constants. 
Taking into consideration also attractive possibilities for particle physics beyond the Standard Model 
related to supersymmetric grand unified theory (SUSY GUT),   
one can consider that    
at higher energies both ordinary and mirror sectors are presented by identical 
GUTs, e.g. $SU(5)$ and  $SU(5)'$ or $SU(6)$ and $SU(6)'$, 
which in both sectors break down to their Standard Model 
subgroups $SU(3)\times SU(2) \times U(1)$ at the scale $M_G \sim 10^{16}$~GeV. 
However, following refs. \cite{Berezhiani:1995, Akhmedov:1992hh}, 
one can assume that later on the symmetry between two sectors is broken
so that the electroweak symmetry breaking scale $V$ in mirror sector is 
much larger than than in the Standard Model, $v = 174$~GeV. 
Namely, if $V\sim 10^{11}$~GeV,  then the lightest shadow baryons 
should have mass order few PeV. Due to baryon violating GUT gauge bosons, 
they decay with $\tau_{\rm dec} \sim t_U$  
producing  energetic shadow neutrinos which oscillate into active neutrinos 
(with oscillation probabilities $\sim 10^{-10}$ or so) 
transferring their spectrum to the latter.



Let us consider, for simplicity,  a supersymmetric grand unification theory $SU(5) \times SU(5)'$.  
 \footnote{ 
As discussed in ref. \cite{BBDG}, our proposal can be more nicely realized in 
SUSY $SU(6)$ theory \cite{Berezhiani:1989bd} 
which gives natural solution to the so called hierarchy and doublet-triplet splitting problems  
by relating the electroweak symmetry breaking scale to the supersymmetry 
breaking scale $M_S \sim 1$~TeV and naturally explains the fermion mass spectrum. } 
Consider  a situation when in both sectors the GUT symmetries are broken at the scale 
$M_G \simeq 2\times 10^{16}$ GeV,  
$SU(5) \to SU(3)\times SU(2) \times U(1)$ and $SU(5)' \to SU(3)'\times SU(2)' \times U(1)'$. 
Below this scale our sector is represented by MSSM with chiral superfields of 
quarks $q_i = (u,d)_i$, $u^c_i, d^c_i$ and leptons $l_i = (\nu,e)_i$, 
$e^c_i$,  $i=1,2,3$ being family index, 
and two Higgs superfields $h_1,h_2$. As for parallel sector we have 
the similar particle content, quarks $Q_i = (U,D)_i$, $U^c_i, D^c_i$,  
leptons $L_i = (N,E)_i$, 
$E^c_i$,  and two Higgses $H_1,H_2$. 
At the scale $\mu = M_G$ all gauge coupling constants  are equal 
and the Yukawa couplings   have the same pattern for ordinary and shadow 
fermions: 
\be{Yuk} 
(Y^e_{ij} l_i e^c_j h_1 + Y^d_{ij} q_i d^c h_1 + Y^u_{ij} q_i u^c_j h_2) ~ + ~ 
(Y^e_{ij} L_i E^c_j H_1 + Y^d_{ij} Q_i D^c H_1 + Y^u_{ij} Q_i U^c_j H_2)   
\ee
%
As for the neutrinos, their masses emerge from the following $D=5$ operators \cite{Berezhiani:1995}:
\be{nus}  
\frac{\alpha_{ij}}{M} (l h_2)(l h_2) +  \frac{\alpha_{ij}}{M} (L H_2)(L H_2) 
+ \frac{\beta_{ij}}{M} (l h_2)(L H_2) 
\ee 
where $M$ is an effective scale of the order of Planck scale $M_{Pl}$ \cite{Akhmedov:1992hh}. 
The first  two terms give the Majorana masses respectively to ordinary neutrinos 
$\nu_{e,\mu,\tau}$ and their shadow (sterile) partners 
while third term induces the mixing between ordinary and shadow neutrinos.

\begin{figure}[!ht]
 \centering
\includegraphics[scale = .7]{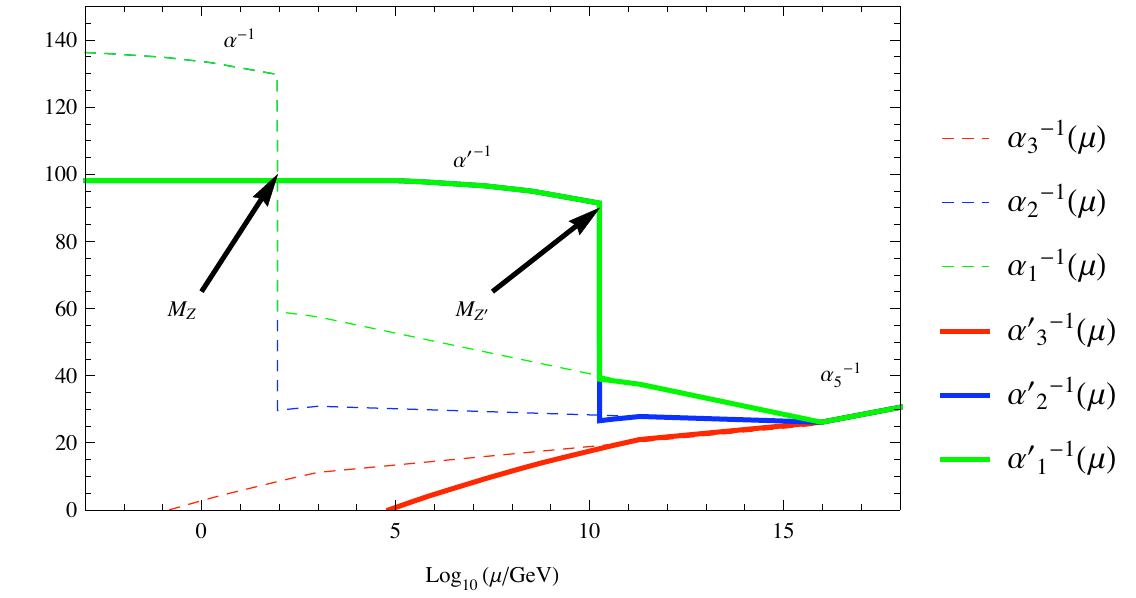}    
\caption{Running of gauge couplings 
below the GUT scale, $\alpha_i = g_i^2/4\pi$ and $\alpha'_i =  g_i^{\prime 2}/4\pi$. }
\label{running}
\end{figure}

Below GUT scale, the gauge coupling constants of two sectors, 
respectively $g_{3,2,1}$ and  $g'_{3,2,1}$ 
evolve down in energies in both sectors in the same way (see Fig. \ref{running}). 
For ordinary sector we follow to the standard paradigm. Supersymmetry is broken 
at the scale $M_S \sim 1$~TeV, the Higgses are not protected anymore to be massless 
and they get VEVs $v_1 = v\cos\beta$, $v_2 = v\sin\beta$, where $v=174$~GeV, 
which induce the electroweak symmetry breaking and generate the fermion masses. 

As for parallel sector, let us assume now that mirror symmetry between
two sectors is spontaneously broken in the following way. 
Imagine that supersymmetry is broken in shadow sector at the scale 
$M'_S \sim 10^{11}$~GeV, supposedly due to non-zero $F$ or $D$ terms of some auxiliary 
fields, so that shadow scalars, including squarks and leptons as well as Higgs doublets 
$H_1$ and $H_2$ acquire soft masses order $M_S$.\footnote{ Interestingly, if 
supersymmetry breaking is transferred to our sector via gravity or 
other Planck scale mediators, this would explain ordinary soft masses order 
$M_S \sim M^{\prime 2}_S/M_{Pl}\sim 1$~TeV. }
Respectively shadow Higgses Get VEVs $V_1 = V\cos\beta'$ 
and $V_2 = V\sin\beta'$, and electroweak symmetry is broken at the scale 
$V\leq M'_S$.  Therefore, the masses of shadow fermions  are rescaled, modulo 
renormalization factors order 1, by a factor $\xi = V/v$ with respect to ordinary fermion masses. 
Namely, taking $V/v = 10^{9}$ and assuming $\tan\beta'=\tan\beta$, 
by the RG running of gauge and Yukawa constants from the GUT scale down in energies, 
one obtains for the shadow electron mass $M_E \simeq 0.4$ PeV, $M_D \simeq 1.1$ PeV 
and $M_U \simeq 1.9$ PeV, against $m_e = 0.5$~MeV, $m_d \simeq 5$~MeV and $m_u\simeq 3$~MeV for ordinary electron and quarks. One the other hand, the RG evolution shows that 
the shadow QCD scale becomes $\Lambda' \sim 100$ TeV  
(c.f. $\Lambda \simeq 200$~MeV in ordinary QCD). Therefore, $\Lambda' \ll M_U, M_D$ 
and the shadow QCD looks like a rescaled version of our QCD without light quarks, 
but only with heavy quarks like $c$ and $b$. Notice also that in shadow sector 
up quark $U$ becomes heavier than the down quark $D$ \cite{BBDG},  and  
$M_{U,D}/\Lambda'$ is of the same 
order as $m_{b,c}/\Lambda$ in ordinary QCD. 
   
The lightest state would appear to be a shadow $\Delta^-$ baryon of spin $3/2$, 
consisting of three down quarks $D$  
and having mass $M_\Delta \approx 3 M_D = 3.3$~ PeV.   
All states, containing up quark $U$, will be unstable against weak decays, 
$U \to D \bar{E} N$. 
As for mesons, 
the lightest pseudoscalar is shadow neutral pion $\pi^0$ consisting of  $D \bar{D}$, 
with mass $M_0 \approx 2 M_D = 2.2$~PeV, while the lightest vector meson 
$\rho^0(D \bar{D})$  is slightly heavier than $\pi^0$. 
Charged Pion $\pi^-$  as well as $\rho^-$-meson  consisting of $D\bar U$
will have mass $M \simeq  M_U + M_D = 3$~PeV,  with $\rho^-$ a bit heavier than $\pi^-$. 
All pseudoscalar  and vector mesons have excited states with mass gap order 
$\Lambda'$ between the levels, just like $c\bar c$ or $b\bar b$ states in ordinary QCD. 

Now we come to the role of baryon violation and proton decay which is fundamental 
prediction of the GUTs. The heavy gauge bosons of $SU(5)$ 
with baryon violating couplings between quarks and leptons induce the decay of 
the lightest ordinary baryons (proton, or neutron bound in nuclei),   
with lifetime $\tau_p \sim M_G^4(\alpha_5^2 m_p^5)^{-1}\sim 10^{31}$~Gyr or so, 
 where  $\alpha_5$ is $SU(5)$ coupling constant at the  GUT scale 
 $M_G \sim 2\times 10^{16}$ GeV \cite{Nath:2006ut}.

In the shadow sector, the similar couplings of GUT gauge bosons should 
destabilize the shadow baryon. 
However, taking into account that shadow $\Delta$ baryon 
is much heavier than the ordinary proton,  $M_\Delta/ m_p \sim 10^6$, 
its lifetime must be about 30 orders of magnitude smaller than the proton lifetime. 
Hence we get $\tau_\Delta \sim M_G^4(\alpha_5^2 m_\Delta^5)^{-1}\sim 10$~Gyr or so, 
with lifetime comparable to the age of the Universe $t_U = 14$ Gyr.   

\begin{figure}[!ht]
 \centering
 \subfigure[]{\includegraphics[scale = 0.3]{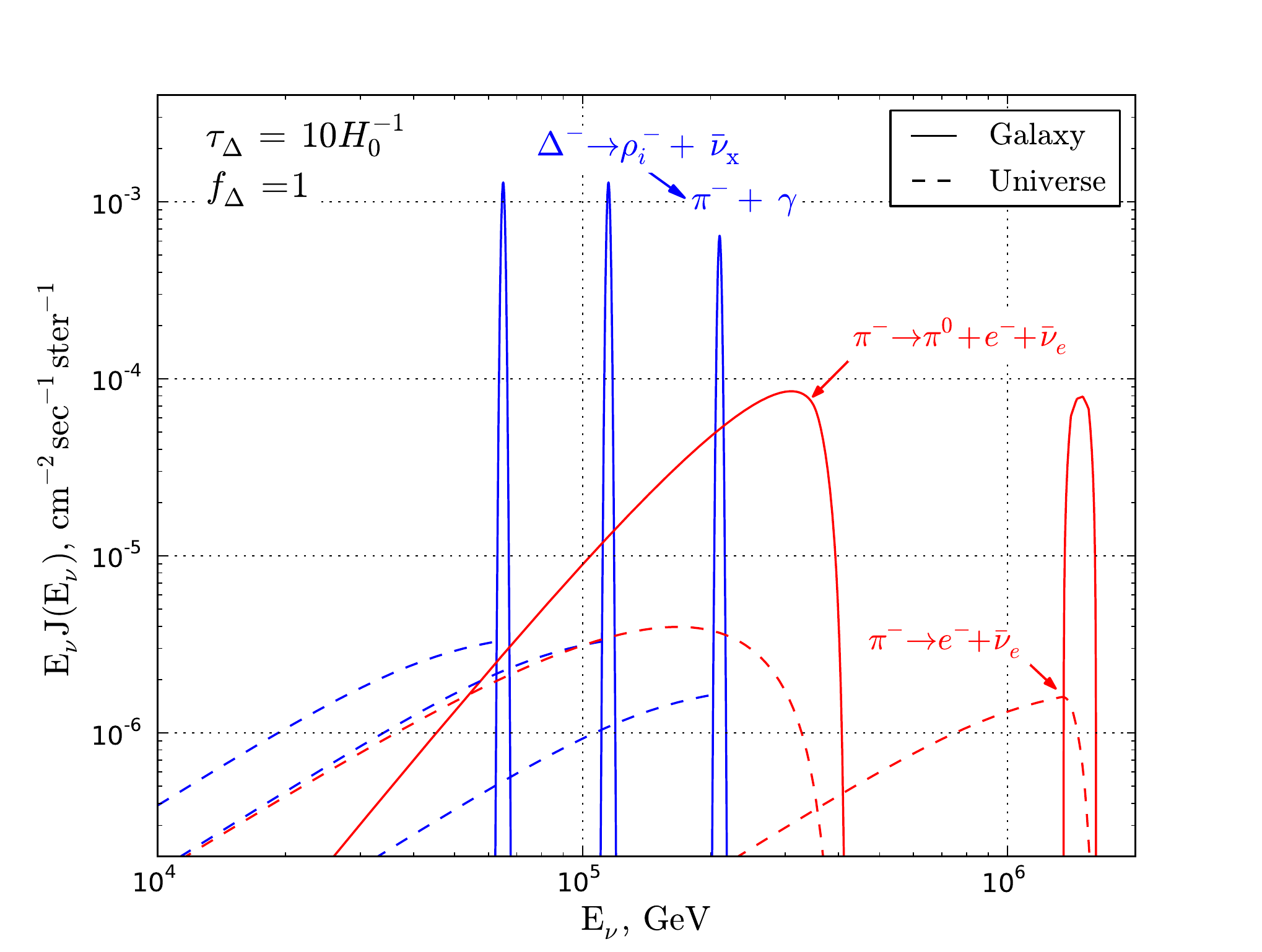}}
 \hspace{5mm}
 \subfigure[]{\includegraphics[scale = 0.3]{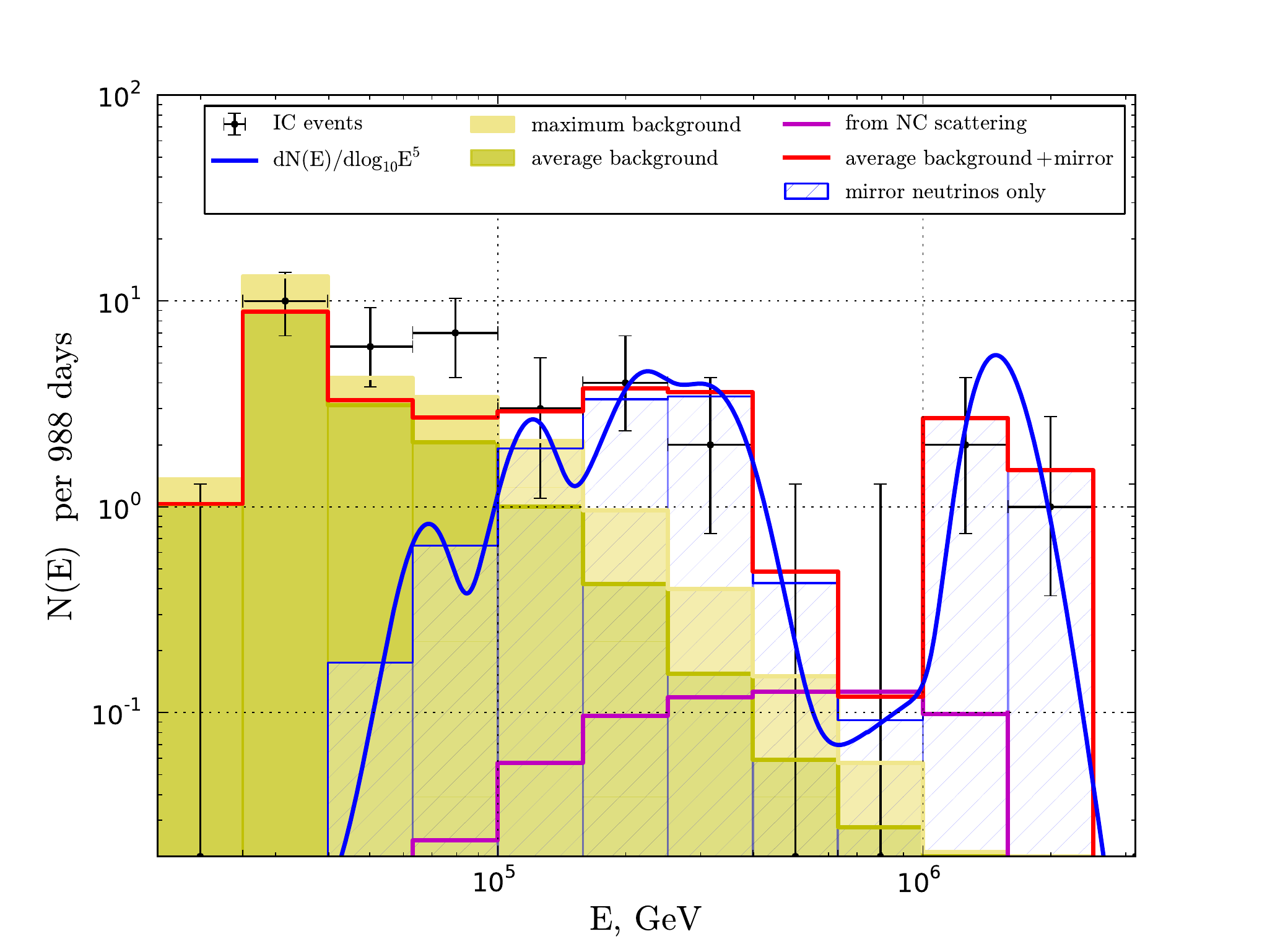}}
 \caption{(a) Shadow neutrino spectrum produced by $\Delta$-barion decay and 
subsequent decay of shadow pions 
(b) Spectrum of VHE neutrino events as predicted in our model. }
\label{spectrum} 
\end{figure}

The principal decay mode of $\Delta$ baryon is in vector mesons, 
$\Delta^- \to \rho^-_{a} + \bar{N}_x$, where  
the neutrino state $N_x$ is generically a superposition of shadow neutrino flavor eigenstates. 
However, taking into account the active sterile mixing $\sim \beta v/\alpha V$ 
due to operators (\ref{nus}), it contains also small admixture of ordinary neutrinos. 
Each decay produces monoenergetic neutrinos $N_x$ with energies 
$E_a = \frac12 M_\Delta (1 - M_a^2/M_\Delta^2)$, where $M_0$ is the mass of 
$\rho^-$ meson and $M_{1,2,...}$ are the masses of its excitations. 
In Fig. 2(a) the spectrum of 
neutrinos produced by decay of galactic dark matter is shown by sharp peaks 
(solid blue) for $M_0, M_1, M_2$ respectively being  $3.0, 3.1$ and 3.2~PeV. 
Due to close degeneracy between the masses of $\Delta$-baryon and 
$\rho^-_a$ mesons, the neutrino energies are $E_a \ll 1$ PeV while   
the most of initial energy $=M_\Delta$ 
is taken away by  mesons $\rho^-_a$. 

Vector mesons promptly decay into mirror pion and photon, 
$\rho^-_a \to  \pi^- + \gamma'$, and subsequent decay of the pion produces the neutrino once 
again (solid red curves in Fig. 2(a)). 
Shadow $\pi^-$ has two decay modes,  two body $\pi^{-}  \to E \bar{N}_E$ and 
three body $\pi^{-}  \to \pi^0 E \bar{N}_E$. Interestingly, their branching ratios 
are comparable which fact is intimately related to the value $\Lambda' \sim 100$~TeV \cite{BBDG}. 
Two body decay produces neutrinos with a narrow energy spectrum concentrated 
around $M_\Delta/2 \simeq 1.6$ PeV, while the three body decay, due to smaller phase space, 
produces less energetic neutrinos with a wide spectrum extending up to the value 
$E_{\rm max} = M_- - M_0 - M_E \simeq M_U - M_D - M_E \simeq 0.4$ PeV. 

Fig. 2(a) shows the final spectrum including the neutrinos produced by the decay of 
dark matter in the galactic halo, and extragalactic neutrinos produced 
by the decay of cosmological dark matter at large redshifts. The fraction of extragalactic 
neutrinos strongly depends on the decay time $\tau_\Delta$ 
(in Fig. 2 we take $\tau_\Delta = 10 t_U$). 
Fig. 2(b) shows how such a spectrum will be seen by the IceCube provided that 
active-shadow neutrino mixing angles are order $10^{-5}$. 
Here the effective areas for the neutrino detection by IceCube \cite{IC} 
and characteristic  error bars in estimation 
 of neutrino energies (of about 13$\%$) are taken into account. 
Needless to say, that the obtained spectrum of events look very much like the spectrum 
observed by the IceCube \cite{IC}. 
The validity of our model will be tested with  increasing statistics by the IceCube collaboration.

\end{document}